\documentclass[10pt,document,superscriptaddress,amsmath,amssymb,aps,pra,floatfix,nofootinbib]{revtex4-2}
\usepackage[T1]{fontenc}
\usepackage{graphicx}
\usepackage{epstopdf}
\usepackage{float}
\usepackage{physics}
\usepackage{amsmath, amssymb}
\usepackage{amsthm}
\usepackage{dsfont}
\usepackage{tabularx}
\usepackage{units}
\usepackage[dvipsnames]{xcolor}
\usepackage{color,soul}
\usepackage{csquotes}
\usepackage{academicons}
\usepackage[section]{placeins}
\usepackage[inkscapelatex=false]{svg}
\usepackage{mathtools}
\MakeOuterQuote{"}

\definecolor{myblue}{named}{NavyBlue}
\definecolor{mygreen}{RGB}{13,255,13}
\usepackage[colorlinks=true,citecolor=myblue,linkcolor=myblue,urlcolor=myblue]{hyperref}
\usepackage[english, capitalise]{cleveref} 
\theoremstyle{definition}
\newtheorem{defn}{Definition}

\theoremstyle{plain}

\theoremstyle{plain}
\newtheorem{lem}[defn]{Lemma}

\theoremstyle{plain}

\theoremstyle{definition}

\theoremstyle{remark}

\begin{document}
	
	%\title{Post-Selection Key Rates for Actively Modulated Single Photon BB84}
    \title{Experimental quantum cryptography with single photons and imperfect devices}
    \author{Aodhán Corrigan $^\clubsuit$}
    \email{aodhan.corrigan@uwaterloo.ca}
	\affiliation{Institute for Quantum Computing and Department of Physics and Astronomy, University of Waterloo, Waterloo, Ontario N2L 3G1, Canada}
    \author{Koray Kaymazlar $^\clubsuit$}
    \email{koray.kaymazlar@tu-berlin.de}
        \affiliation{Department for Quantum Technology, University of Münster, 48149 Münster, Germany}
        \affiliation{Institute for Physics and Astronomy, Technical University of Berlin, 10623 Berlin, Germany}

    \author{Zhiyao Wang}
	\affiliation{Institute for Quantum Computing and Department of Physics and Astronomy, University of Waterloo, Waterloo, Ontario N2L 3G1, Canada}

    \author{Lucas Rickert}
    \altaffiliation{Present address: Toshiba Europe Ltd., 208~Cambridge Science Park, Milton Road, Cambridge, CB4~0GZ, United Kingdom}
    \affiliation{Institute for Physics and Astronomy, Technical University of Berlin, 10623 Berlin, Germany}
    
    \author{Daniel Vajner}
    \altaffiliation{Present address: Advanced Quantum Light Sources GmbH, Guerickestraße 12, D-10587 Berlin}
    \affiliation{Institute for Physics and Astronomy, Technical University of Berlin, 10623 Berlin, Germany}

    \author{Martin von Helversen}
    %\email{helversen@physik.tu-berlin.de}
    \affiliation{Institute for Physics and Astronomy, Technical University of Berlin, 10623 Berlin, Germany}

    \author{Hanqing~Liu}
    \author{Shulun~Li}
    \author{Haiqiao~Ni}
    \author{Zhichuan~Niu}
    \affiliation{Key Laboratory of Optoelectronic Materials and Devices, Institute of Semiconductors, Chinese Academy of Sciences, Beijing, 100083, China}
    \affiliation{Center of Materials Science and Optoelectronics Engineering, University of Chinese Academy of Sciences, Beijing, 100049, China}

    \author{Devashish Tupkary}
    \affiliation{Institute for Quantum Computing and Department of Physics and Astronomy, University of Waterloo, Waterloo, Ontario N2L 3G1, Canada}
    
    \author{Tobias Heindel}
    %\email{tobias.heindel@tu-berlin.de}
    % \affiliation{Institute for Physics and Astronomy, Technical University of Berlin, 10623 Berlin, Germany}
    \affiliation{Department for Quantum Technology, University of Münster, 48149 Münster, Germany}
    
	\date{\today}
	\begin{abstract}
    Quantum key distribution (QKD) allows for provably secure key distribution between two trusted parties. Because the security and performance of QKD protocols rely on devices that behave according to specific assumptions, idealized or inaccurate assumptions about device behavior can introduce security loopholes. Real devices can never be perfectly characterized, and their performance metrics are always subject to certain error margins, which must be accounted for in a rigorous theoretical analysis. Only recently have rigorous finite-size results allowed for imperfect characterizations of devices (where device parameter  have uncertainty margins) - an advance yet to be considered in experimental implementations of the BB84 protocol. In this work, we prove the security and analyze the performance of an implementation of the BB84 protocol using single photons generated by a semiconductor quantum dot light source in combination with dynamic polarization-state encoding. We consider the presence of incompletely characterized devices by accounting for imperfections in the single-photon source (in terms of finite $g^{(2)}(0)$) as well as the receiver (non-ideal beam-splitters, finite detector efficiencies, and dark counts), all with error margins. The resulting protocol implementation shows competitive performance, paving the way towards practical and loop-hole free implementations of QKD.
	\end{abstract}

	\maketitle
    \def\thefootnote{$\clubsuit$}\footnotetext{These authors contributed equally to this work.}
    \renewcommand{\thefootnote}{\arabic{footnote}}
    \setcounter{footnote}{0}
    \renewcommand{\thefootnote}{\arabic{footnote}}
    \renewcommand{\thefootnote}{\arabic{footnote}}
    \setcounter{footnote}{0}
    
\section{Introduction}\label{sec:Intro}
Quantum Key Distribution (QKD) allows for information-theoretic secure key distribution, by exploiting laws of quantum physics rather than computational assumptions. Major efforts are underway to implement practical QKD devices worldwide, and similarly, there are pushes to prove security for these devices in practical settings while reducing the assumptions about physical hardware and accounting for imperfections \cite{michaelPassiveEUR, nahar2025imperfectdetectorsadversarialtasks, nahar2026proof, kamin2025renyisecurityframeworkcoherent,  curráslorenzo2025securityquantumkeydistribution, Curr_s_Lorenzo_2025, curraslorenzo2026rigorousphaseerrorestimationsecurityframework, Xu_2020,https://doi.org/10.1002/qute.202300380, Li_2025}. 
The gap between the idealized assumptions in QKD security analyses and experimental implementations has been a long standing problem in ensuring implementation security, with several works proving security in the presence of (imperfectly characterized) imperfections such as polarization and intensity imperfections \cite{curráslorenzo2025securityquantumkeydistribution, curraslorenzo2026rigorousphaseerrorestimationsecurityframework}, and in the presence of detector imperfections (such as dark count events, unequal efficiencies and memory effects), \cite{michaelPassiveEUR, https://doi.org/10.1002/qute.202300380, nahar2025imperfectdetectorsadversarialtasks, nahar2026proof}. 
When analyzing the security and performance of experimental implementations of QKD, device imperfections in the protocols are typically unaccounted for, or idealized claims are made in which devices are assumed to be perfectly characterized, which in turn opens the door to potential side channels through which an adversary can gain information about the shared secret key. In this work, we consider both source-side (non-negligible multiphoton probability) as well as detector-side (beamsplitting and detector efficiency mismatches as well as unequal dark count rates) imperfections that are characterized within some range and prove security in the Entropic Uncertainty Relation (EUR) framework \cite{tomamichel2011uncertainty, Tomamichel2017largelyself}, building upon recent theoretical work by Ref. \cite{michaelPassiveEUR}. We implement the protocol using a state-of-the-art semiconductor quantum dot (QD) light source and employ high speed dynamic polarization encoding without decoys, with a fully passive Bob constructed from a Superconducting Nanowire Single Photon Detector (SNSPD) based four-state analyzer. 

Attenuated lasers are typically used in implementing QKD protocols due to their reliability and commercial availability. However, lasers emit coherent states which have significant multiphoton components, which makes them inherently vulnerable to photon number splitting attacks \cite{PNS, Calsamiglia_2001}. Accounting for this requires decoy states, which introduces extra experimental complexity because of the need for an additional RNG source and careful modulation of source intensities \cite{decoyHwang, decoyHKLo, decoyWang}. In contrast, the use of QDs for QKD is motivated by their desirable quantum-optical properties as they are closer to true single photon sources. Notably, they allow for efficient and controllable sub-Poissonian light generation due to their high multiphoton suppression, ability to generate indistinguishable photons and high relative brightness \cite{ding2023highefficiencysinglephotonsourcelosstolerant, Tomm_2021, Somaschi_2016, Schweickert_2018, Rickert_2024, Rickert_2025}. True single photon sources capable of generating so-called "flying qubits" are desirable not only for quantum cryptographic applications, but also in other quantum computing applications where true qubits are needed \cite{doi:10.1126/science.1142892}. Furthermore, our dynamic polarization modulation circumvents the need for having multiple indistinguishable light sources set to different polarizations during state preparation, eliminating a possible side channel attack arising from intensity imperfections in different states \cite{pereira2022modifiedbb84quantumkey, kamin2025renyisecurityframeworkcoherent}.

Thus far, even though there have been many implementations of cryptographic protocols with QD sources, an implementation which considers rigorous device characterization in its security proof is not present in the literature. Recent theoretical analyses, which we make use of in our analysis, allow for much more relaxed characterization of the optical detector modules used in the protocol (See Sec. \ref{sec:proofOverview} for discussions). We demonstrate competitive key rates with our experimental system in the presence of such device imperfections, which is an important step in the direction of actual QKD implementations and demonstrates the viability of QD sources for cryptographic applications.  

The paper is organized as follows: In Section \ref{sec:Results} we describe our experimental implementation as well as measured values, compute key rates for our protocol and discuss the limitations of our analysis. In Section \ref{sec:Conclusion} we provide some concluding remarks. Finally in Appendix \ref{AppendixProof} we prove security for our implementation, and in Appendix \ref{appendix:blinking} we explain how we can correct for blinking in the source $g^{(2)}(\tau=0)$ measurement.
\section{Implementation and Results}\label{sec:Results}
\subsection{Experimental Setup}
\begin{figure}[h]
    \centering
    \includegraphics[width=1\linewidth]{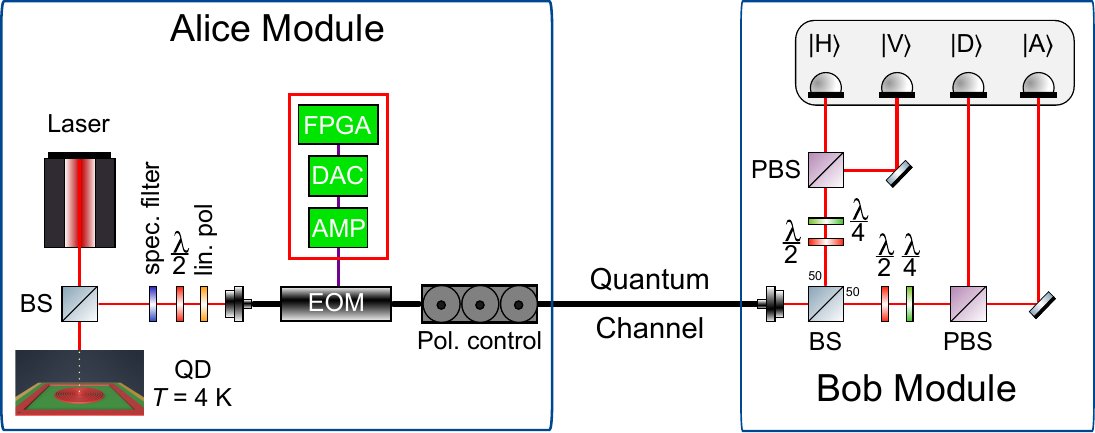}
    \caption{Experimental setup for our BB84 implementation. Alice prepares single photons using a deterministic single photon source comprised of a single QD embedded in a photonic microcavity and dynamically selects the polarization using an EOM. These are sent to Bob's passive four-state polarization analyzer, comprised of SNSPD based detectors.}
    \label{fig:lab_setup}
\end{figure}
We implement the BB84 protocol using the experimental setup shown in Fig. \ref{fig:lab_setup}. In each round, Alice prepares a single-photon pulse using a QD-device and encodes the classical information 0 or 1 via its polarization either in the $\mathcal{Z}$ (states $\ket{H}$ and $\ket{V}$) or the $\mathcal{X}$ (states $\ket{D}$ and $\ket{A}$) basis. State encoding is performed via an electro-optical modulator (EOM) that prepares the four BB84 states $\ket{\xi}\in\{\ket{H},\ket{V},\ket{D},\ket{A}\}$. The prepared flying qubits propagate through a free space quantum channel over to Bob, who then detects the states using a passive four-state polarization analyzer. Basis and sifting announcements are performed through a classical data link. We expand on the experimental details in the following subsections, while mathematical formalizations of the protocol as well as details on the classical post-processing can be found in Appendix \ref{AppendixProof}.

\subsubsection{Single Photon Generation}
\begin{figure}
    \centering
    \includegraphics[width=1\linewidth]{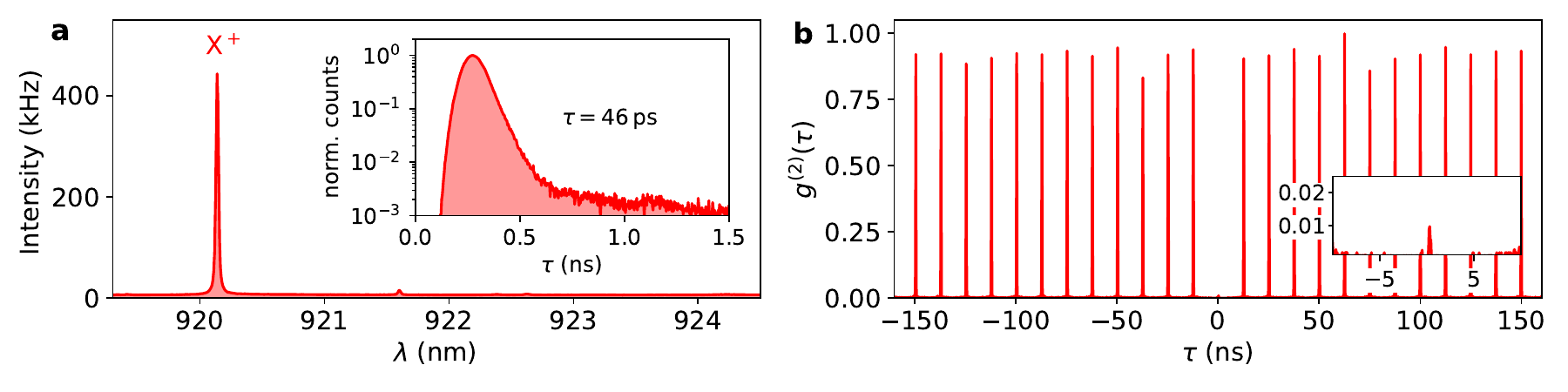}
    \caption{a) Emission spectrum of the QD microcavity device under quasi-resonant (p-shell) excitation. The strongly Purcell-enhanced trion state ($X^+$) at 920.1 nm exhibits a ultra-fast decay time $<50$\,ps, as depicted in the time-resolved measurement in the inset. b) Second-order photon correlation histogram measured in Hanbury-Brown and Twiss configuration. The strong multiphoton suppression of $g^{(2)}(\tau=0) = (1.7\pm0.5)\cdot 10^{-2}$ demonstrates sub-Poissonian emission statistics.}
    \label{fig:dotPerformance}
\end{figure}

For the generation of single photons Alice uses a deterministic single-photon source (SPS) based on a pre-selected InAs QD embedded in a hybrid circular Bragg grating cavity featuring pronounced Purcell enhancement \cite{Rickert_2024}. The source emitting at about 920 nm is operated in a cryogenic environment at 4 K and is excited quasi-resonantly using a pulsed laser at 896 nm wavelength, yielding the measured emission spectrum and photon-autocorrelation histogram $g^{(2)}(\tau)$ depicted in Fig. \ref{fig:dotPerformance}(a) and (b), respectively. We obtain an integrated multiphoton suppression value $g^{(2)}(\tau=0)$ of ($0.017\pm0.005)$ after blinking correction (see Sec.~\ref{sec:devices}). Noteworthy, the selected excitation scheme results in vanishing photon-number coherence \cite{vajner2026coinflip}, as required for the protocol implementation discussed in this work. 

\subsubsection{Dynamic Polarization Modulation}
\begin{figure}
    \centering
    \includegraphics[width=1\linewidth]{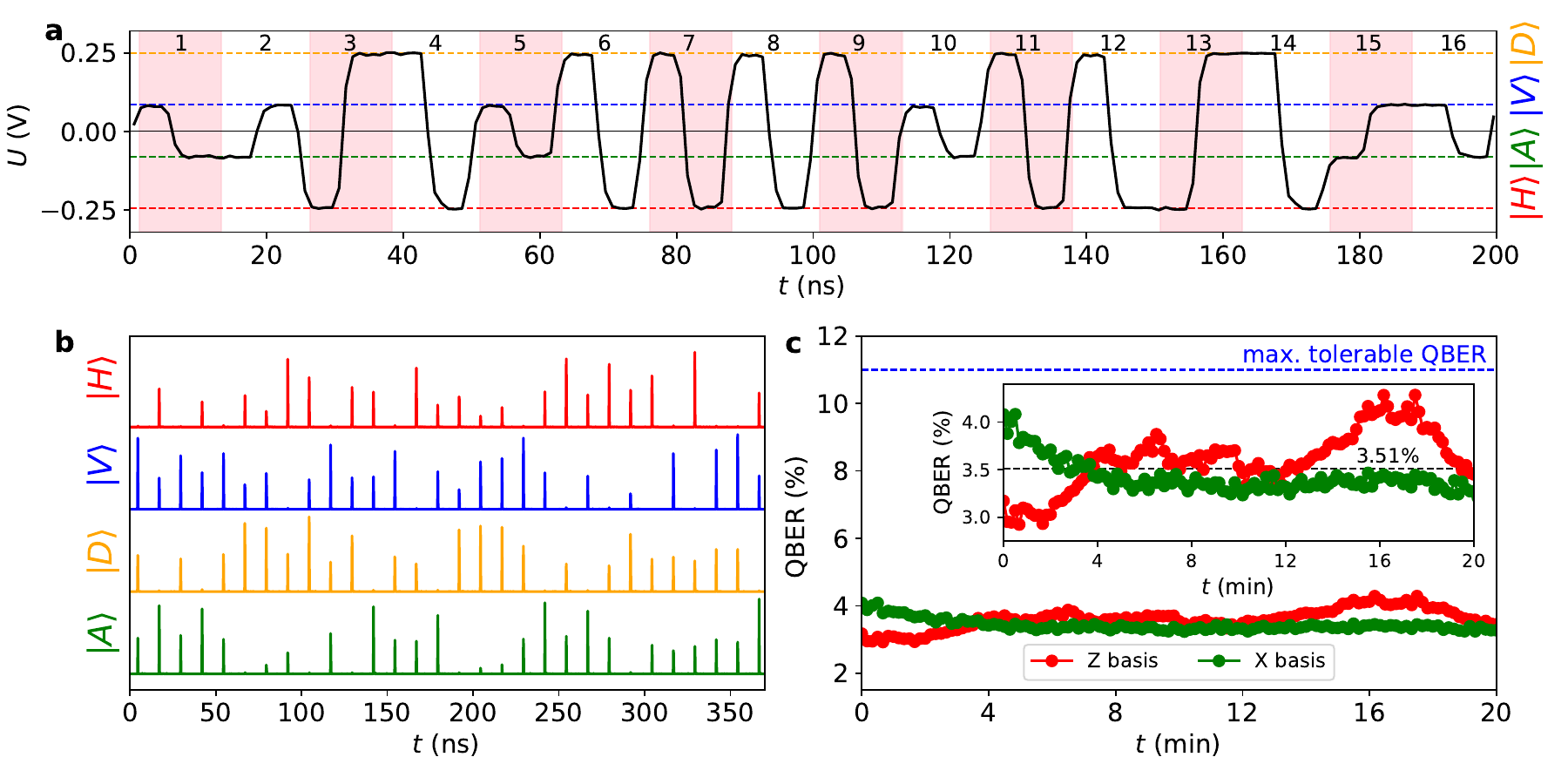}
    \caption{ (a) Example of the employed random sequence for a 200~ns long excerpt containing 16 bits of given voltage level $U$ in the four polarization states, followed immediately by the binary switching in each bit as inspired by the Manchester encoding. (b)  Sample of time-tags averaged over multiple QRNG sequence loops, showing high polarization suppression in all four bases over $\sim$400\,ns of the QRNG sequence. (c) Maximum tolerable QBER of 11\% for BB84 QKD and achieved QBER for $\mathcal{Z}$ and $\mathcal{X}$ bases over the course of the experiment (i.e. 20 mins), with a time average of 3.51\%.}
    \label{fig:myPair}
\end{figure}

Alice dynamically and randomly prepares the polarization states to be sent to Bob using a customized fiber-coupled EOM controlled by a custom-built arbitrary waveform generator consisting of a field programmable gate array (FPGA) evolution board and a 16-bit digital to analog converter (DAC). The FPGA is fed with a pre-compiled $10^5$ quaternary digit random sequence sourced from the Australian National University QRNG API \cite{ANU_QRNG}, which generates random strings by measuring vacuum fluctuations. Quaternary bits are needed to choose one of Alice's four states. In each clock cycle, the main circuit in the FPGA reads a two-bit number from the block random access memory and generates a 16-bit digital output to drive the DAC. The analog signal from the DAC is amplified via an RF amplifier before reaching the EOM. Figure \ref{fig:myPair}(b) displays exemplary detection time traces recorded on Bob's side at the four detection channels. To avoid voltage level drifts on short time scales, caused by the AC-coupled DAC output, we implemented a solution inspired by Manchester coding. Switching from each prepared voltage level $x$V to its negative value $-x$V, we enforce a vanishing average voltage in random voltage sequences (Fig. \ref{fig:myPair}(a)). While this significantly reduces the achievable QBER, it requires a more precise synchronization with the short single-photon pulses as the FPGA-internal clock's frequency $f_{\textrm{FPGA}}=160\,$MHz generated via a phase locked-loop circuit is double the repetition rate of the excitation laser ($f_{\textrm{laser}}=80\,$MHz). Using this approach, we observe a net decrease in QBER and improved temporal stability, i.e. reduced QBER fluctuations over time, with an average QBER across the $\mathcal{Z}$ and $\mathcal{X}$ bases of 3.51\% for 20\,mim accumulation time (see Fig.  \ref{fig:myPair}(c)). Note that in comparison to periodic four-state switching sequences (wherein the waveform fed to the EOM is repeated after four clock cycles, for example $\circlearrowright H\to V\to D\to A\circlearrowright$, typically used for benchmarking), random sequences, as required in QKD for security, cause increased voltage level jitter, resulting in higher average QBERs in random sequences. 

\subsubsection{State detection}
Polarization encoded photons are detected in a four-state polarization analyzer comprised of a 50:50 beamsplitter at the input of the free space channel which steers photons towards either the $\mathcal{X}$ or $\mathcal{Z}$ basis detectors completely passively. Each basis detection is comprised of a polarizing beamsplitter polarized in the respective basis, followed by two superconducting nanowire single photon detectors (SNSPDs) optimized for the different orthogonal states via polarization control paddles at the input. Bob receives a conclusive measurement if he records a single click event in one detector, and associatess the thisparticular outcome withwith a basis decision. We emphasize that Bob does not actively choose which basis to measure in, and that this basis decision is implemented passively and depends on the detected events being conclusive. Inconclusive events, consisting of no- or multiclick events, are recorded as well but are not used for key generation.

\subsection{Experimental Results}
In the actual QKD experiment, Alice prepares single photons with random polarization $\xi$ in each round and according to her random sequence, which is repeated over the course of the protocol to obtain sufficient statistics. This prepare-and-measure implementation runs at the 80 MHz optical clock rate for an accumulation time of 20\,min, resulting in $1.1\cdot 10^{11}$ signal events sent by Alice. Synchronization of the state preparation and detection events is achieved by a classical communication channel linking Alice's excitation laser clock and the polarization encoder to Bob's detector time tags. A random subset of the protocol's rounds are used to determine the exact temporal offset during the protocol run by shifting the data until the highest measured correlation is achieved as a part of the classical post-processing. Further post-processing is done in order to only record measurement outcomes where Alice and Bob's basis decisions agree (i.e. rounds in which Bob received a conclusive outcome in the same basis Alice announced for encoding). The kept $\mathcal{Z}$ basis rounds are set aside for key generation, while the $\mathcal{X}$ basis rounds are used for testing and parameter estimation.

A key figure of merit is the QBER in $\mathcal{Z}$- and $\mathcal{X}$-basis, which are computed as the ratio of incorrect measurements in the $\mathcal{Z}$ and $\mathcal{X}$ bases ($n_{\neq, \mathcal{Z}}$ and $n_{\neq, \mathcal{X}}$ respectively) and all signals in that basis that resulted in conclusive outcomes ($n_{\mathcal{Z}}$ and $n_{\mathcal{X}}$, respectively). Conclusive signals are composed of rounds in a basis following sifting, which is the process of removing rounds where either Alice's encoding basis and Bob's measurement basis do not match, or rounds in which Bob received no detections. Note that unlike other EUR based security proofs (See, for example, Ref. \cite[Section 3.2.4.]{tupkary2025qkdsecurityproofsdecoystate}) we do not assign double click events to randomly chosen single click events, instead recording the amount (in addition to cross-basis clicks) as the quantity $n_\mathrm{mc}$. Therefore, double click events do not contribue to the QBER.
Explicitly, 
\begin{align}
    e_{\mathcal{Z}} = \frac{n_{\neq,\mathcal{Z}}}{n_{\mathcal{Z}}},\ \text{and}\ e_{\mathcal{X}} = \frac{n_{\neq,\mathcal{X}}}{n_{\mathcal{X}}}.
\end{align}
Overall, we observe approximately $7.66\cdot 10^7$ detection events stemming from $1.07\cdot 10^{11}$ attempted communication rounds over the course of the 20 minute experiment, resulting in a basis-averaged QBER of 3.51\%. These values are listed in Table \ref{tab:values_pair}, and we discuss the device characterization in subsection \ref{sec:devices}. 
\begin{table}[H]
\centering
\begin{minipage}[t]{0.4\linewidth}
\centering
\fbox{%
\begin{tabular}{c|c|c}
\textbf{Device Parameters} & \textbf{Value} & \textbf{Uncertainty} \\
\hline
Source Mean Photon Number $\mu$ & 0.005 & 10\%\\
Source Autocorrelation $g^{(2)}(0)$ & 0.017 & 30\%\\
Beamsplitter Ratio $s$ & 0.496 & 2.5\% \\
Detector Relative Efficiencies $\vec{\eta}$ & (1,1,1,1) & 2.5\%\\
Dark Count Probability $p_\mathrm{dc}$ & $10^{-7}$ & 50\%\\
\end{tabular}%
}
% optional: \label{tab:qber}
\end{minipage}%
\hspace{0.1\linewidth}%
\begin{minipage}[t]{0.4\linewidth}
\centering
\fbox{%
\begin{tabular}{c|c}
\textbf{Recorded Events/Measurements} & \textbf{Value} \\
\hline
Attempted Rounds $n$ & $107364\cdot 10^6$\\
Detected Events $n_\mathrm{det}$ & $76627590$\\
Multiclick Events $n_\mathrm{mc}$ & 20 \\ 
Sifted $Z$ basis events $n_\mathcal{Z}$& 19023775 \\ 
Sifted $X$ basis events $n_\mathcal{X}$& 19315816 \\ 
Time Averaged $\mathcal{Z}$ QBER & 3.61\% \\
Time Averaged $\mathcal{X}$ QBER & 3.41\% \\ 
\end{tabular}%
}
% optional: \label{tab:values}
\end{minipage}
\caption{Values recorded during experiment. (Left) Device parameters used in simulation and key rate calculation. (Right) Values actually observed during run of experiment.}
\label{tab:values_pair}
\end{table}
While we end our discussion of our experimental implementation of the protocol here, an actual implementation of QKD would then continue its classical postprocessing in order to generate the final secret key. Next, Alice and Bob implement the variable length decision, which determines the final key length they will hash to, and in turn fixes the amount of communication rounds they will perform for error correction and verification. Afterwards error correction is implemented by exchanging messages over the authenticated classical channel \cite{tupkary2026authenticationsecurityproofsquantum}, a process which leaks $\lambda_\mathrm{EC}$ bits of information. Then error verification is performed using a randomly chosen 2-universal hash function and exchanging values over the classical channel, which succeeds with probability $1-\varepsilon_\mathrm{EV}$ and leaks $\log(2/\varepsilon_\mathrm{EV})$ bits, and in turn indicates whether error correction was successful. Finally, privacy amplification is implemented using another randomly chosen 2-universal hash function, mapping the corrected bit strings to shorter ones, with the length being determined by the variable length decision. Classical communication in this step leaks a further $2\log(1/\varepsilon_\mathrm{PA})$, with the step succeeding with probability $1-\varepsilon_\mathrm{PA}$. For a rigorous description of these sub-protocols, we refer the reader to Ref. \cite[Section 4]{tupkary2026rigorouscompletesecurityproof}.

\subsubsection{Key Rate Performance}
\begin{figure}[H]
    \centering
    \includegraphics[width=0.65\linewidth]{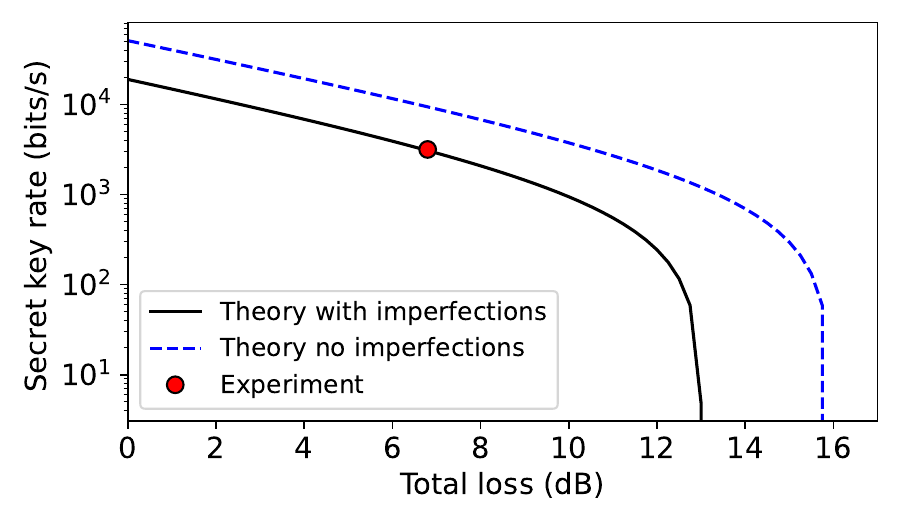}
    \caption{Simulated key rate plot for experimental parameters, as well as key rate computed from experimental run. Simulation is performed by using error rate observed in experiment, but changing loss. Computed with sampling parameters $\varepsilon_\mathrm{PNE}=\varepsilon_{a}=\varepsilon_{b}=\varepsilon_{0}=\varepsilon_{\mathrm{AT},Z}=10^{-9}$ (See Appendix \ref{AppendixProof} for details), postprocessing parameters $\varepsilon_{\mathrm{PA}}=\varepsilon_{\mathrm{EV}}=10^{-9}$, and probability of $\mathcal{Z}$ rounds used for $e_\mathcal{Z}$ estimation of 1\%. Horizontal axis begins at 0, assuming lossless components. Note accounting for imperfections drastically reduces key rate by nearly an order of magnitude, although further refinements of proof techniques will likely be able to decrease this impact.
    % Using an attenuation rate of $2 \mathrm{dB}/\mathrm{km}$ for generic single mode fiber optics \cite{heitmann1990attenuation}, the horizontal axis is equivalent to 7.5 km of fiber distance. Similarly, for free space transmission with clear visibility at 920 nm yielding approximately 0.40 dB/km \cite{green2019optical}, the horizontal axis would correspond to roughly 37.5 km of free space distance. 
    }
    \label{fig:KeyRate}
\end{figure}

%which collectively denote by $\vec{F}^\mathrm{obs}$, from measurement rounds that are announced.
During the performance of the protocol Alice and Bob generate an array of observed data statistics, which we collectively denote as $\vec{n}^{\mathrm{obs}}$. This includes all of the click patterns (including no-clicks) observed during the run of the experiment, which we can sift through to compute the number of conclusive detections in each basis and the number of multiclick events. From these components we can also compute the the $\mathcal{X}$ basis QBER and a sample of the $\mathcal{Z}$ basis QBER from some small testing fraction of the $\mathcal{Z}$ basis rounds, with the latter being used to estimate the cost of error correction. The goal of Alice and Bob is to make use of the observations to determine the key length to hash to, a quantity which we use in order to gauge the performance of the QKD protocol.  In the presence of device imperfections, the key length for the protocol is given by 
\begin{align}
    \ell = \max \left\{\mathcal{B}^\mathrm{Single}_{q_Z}\left(\vec{n}^\mathrm{obs}\right)\cdot\left(1-h\left(\mathcal{B}^\mathrm{Error}_{a,\delta,q_Z}\left(\vec{n}^\mathrm{obs}\right)\right)\right) - \lambda_\mathrm{EC}(\vec{n}^\mathrm{obs}) - 2\log(1/2\varepsilon_\mathrm{PA}) - \log(2/\varepsilon_\mathrm{EV}),0 \right\}
\end{align}
where $\mathcal{B}_{q_Z}^\mathrm{Single}\left(\vec{n}^\mathrm{obs}\right)$ is an estimate on the amount of single photon contributions to the key generation rounds computed as a function of the observations and Alice's source properties, $\mathcal{B}^\mathrm{Error}_{a,\delta,q_Z}(\vec{n}^\mathrm{obs})$ is an estimate of the so-called phase error rate (which can be thought of the error rate in $\mathcal{Z}$ basis rounds had they been measured in the conjugate $\mathcal{X}$ basis) computed from observations, Alice's source properties as well as parameters $a,\delta$ and $q_Z$ which are related to device properties and can be explicitly computed from the experimental device characterizations and specifications (See Eq.'s D40, D41 and E2 respectively in Ref. \cite{michaelPassiveEUR}), $\lambda_\mathrm{EC}$ is the number of bits used in the classical error correction protocol \footnote{To be precise, it is the logarithm (to the base $2$) of the number of possible transcripts of the error correction protocol. In the event that Alice sends a fixed number of bits, this corresponds to the number of bits.}, and $\varepsilon_{\mathrm{EV}}$ and $\varepsilon_\mathrm{PA}$ are the failure probability of error verification and privacy amplification, respectively. For more details on the derivation and exact expression, we refer the reader to Appendix \ref{AppendixProof}. These statistical bounds are computed in terms of the quantities observed during the run of the protocol (Table \ref{tab:values_pair}), and so allow us to compute the key length that would be hashed to at the end of the protocol.  We compare this to a simulated channel with the same device parameters while scanning over channel loss, in order to simulate the performance under realistic conditions. The key rate is simulated assuming lossless components in Bob's device\footnote{In proving security, it is possible to "factor out" common loss via noise channel constructions, such that the loss in the optical devices is modeled as channel loss. See, for example, Ref. \cite{nahar2025imperfectdetectorsadversarialtasks} for a rigorous treatment.}, while our experiment incurs attenuation due to the channel and optics involved (roughly $7$ dB). For our experiment, we compute a secure key length of $ 2.1579\cdot 10^6$ bits in the presence of device imperfections and uncertainties, or roughly 250 kilobytes (Fig. \ref{fig:KeyRate}).

We emphasize that although the key rates computed may seem lower than recent analyses \cite{zahidy2024quantum, morrison2023single, yang2024high}, the above calculations explicitly include both device imperfections and uncertainties in the characterization of such imperfections.

\subsection{Overview and Limitations of Security Proof}\label{sec:proofOverview}
When performing security analyses of QKD protocols, device imperfections can be broadly grouped into source and detector categories, whereby the way to deal with these differs based on the chosen proof technique \cite{tupkary2025qkdsecurityproofsdecoystate}. In this work, we make use of a phase-error-based proof technique and therefore focus our discussion on this approach. Source imperfections have also been rigorously considered \cite{tamaki2014loss, curráslorenzo2025securityquantumkeydistribution, curraslorenzo2026rigorousphaseerrorestimationsecurityframework} since imperfect polarization preparation, correlation effects and intensity imperfections can lead to information leakage. Multiple approaches have also been developed for dealing with detector imperfections known only within a range of values, rather than fixed exactly and staying constant for the duration of an experiment, in the finite size regime and against coherent attacks \cite{michaelPassiveEUR, curráslorenzo2025securityquantumkeydistribution, nahar2025imperfectdetectorsadversarialtasks, nahar2026proof, tupkary2024phaseerrorrateestimation}. This is necessary since detector efficiencies and dark count rates can vary during the run of an experiment, most commonly due to temperature fluctuations in the detectors \cite{dennis2012photodetectors}. Importantly, detector-side imperfections also include imperfect characterizations of beamsplitting ratios, which is a core component of the linear-optical BB84 protocol as it implements Bob's measurement basis choice. In particular, detector imperfections in passive detector setups have only recently been rigorously addressed  \cite{michaelPassiveEUR}, though this has not yet been combined with source imperfections\footnote{The combination of source and detector imperfections for passive protocols in the EUR based approach is still open. While theoretically possible (See Section VIIIA of Ref. \cite{michaelPassiveEUR}), the formulation there appears to be a computationally cumbersome optimization problem when using imperfect single photon sources.}. Regardless, we can still consider imperfect characterization of Bob's detector module, something which, to the best of our knowledge, has yet to be done in experimental implementations of the BB84 protocol. We further consider rigorous finite-size security bounds that have recently been developed and shown to perform well in the presence of such imperfections \cite{michaelPassiveEUR}. We present our security proof in full detail in Appendix \ref{AppendixProof}.

 %Importantly as well, no framework to-date is capable of considering Trojan Horse or other types of light injection attacks on quantum dot systems  \cite{pereira2019quantum, pereira2022modifiedbb84quantumkey,pereira2020quantum, Curr_s_Lorenzo_2025, mizutani2021security}.We leave a more rigorous analysis of these attacks on QD systems to future work.  %

\subsection{Characterization of Devices}\label{sec:devices}
Errors or fluctuations in device characterization must be accounted for in proving security. In our analysis, we considered detector-side imperfections in the form of imperfect beamsplitter characterization as well as dark count and efficiency rates that are only known within some range, and source-side imperfections in the form of fluctuating emitter statistics. These values are listed in Table \ref{tab:quantities}.

In order to characterize the detector module, the beamsplitter ratio uncertainty was measured by shining attenuated pulsed optical parametric oscillator (OPO) laser light at 920$\,$nm into the beamsplitter and measuring the intensity of the output on each side (corresponding to the $\mathcal{Z}$ and $\mathcal{X}$ basis outputs). We measured $63\pm1.8\,$µW reflected and $64\pm1.8\,$µW transmitted with a photodiode and the manufacturer stated measurement uncertainties, corresponding to a beamsplitter ratio of $s=\sqrt{r}=0.496\pm0.014$, and $1-s=\sqrt{t} = 0.504\pm 0.014$, corresponding to an uncertainty of 2.5\%. Assuming perfect PBS extinctions, we were then able to compute the relative detector efficiencies. Using unpolarized light (equivalent to polarization-randomized states), we flooded the detectors with an attenuated pulsed laser at 920$\,$nm and adjusted the detector efficiencies until all were equalized to the same count rate with no count fluctuations after temperature stabilization. The efficiencies were adjusted by changing the nanowire bias voltage and the pitch adjusters in the fibre couplers. In order to compute the dark count fluctuations, we blinded the detectors from external light and recorded counts, which fluctuated around $20\pm 10\,$Hz. 

With regards to imperfections on side of the QD source, we corrected for blinking of the investigated QD emitter (see Appendix~\ref{appendix:blinking}) and obtained a $g^{(2)}(0)$-value of $(0.017\pm0.005)$. The mean photon number in the quantum channel is measured as $\mu=(0.0050 \pm 0.0005)$. Polarization states are assumed to be perfect at Alice's output, although we measure these with a polarimeter to a high degree of accuracy. 

 We further assume Alice's lab is completely isolated from Eve's influence, meaning no light injection attacks are possible. Future work may want to study the impact of optical isolators on quantum dot-based systems. Finally, our time-syncing technique uses protocol signals to determine time correlations. To avoid this, synchronized clocks may be used instead.
 
\section{Conclusion and Outlook}\label{sec:Conclusion}
In this work, we made use of recent theoretical bounds to prove security of a QD-based BB84 implementation with imperfectly characterized detectors, as well as source emission statistic imperfections. This analysis addresses important imperfections that comes from realistic devices, and yet we still demonstrate competitive performance. Our experiment made use of dynamic polarization encoding for over 20 minutes of integration time, yielding around $10^{11}$ rounds of the protocol. Given this, we observed a slightly fluctuating average QBER of 3.5\%. Our implementation further exploited the source's low $g^{(2)}(0)$ value, which allows for high multiphoton suppression, to remove the need for decoy-style intensity modulation. Through this, we would have been able to hash to a key length of roughly $2\cdot 10^6$ secure bits. 
As discussed in Sec. \ref{sec:proofOverview}, while our analysis rigorously considers detector-side imperfections, we are unable to account for source imperfections. This means we do not consider flaws in the polarization preparation, though we were able to account for fluctuating photon number emission statistics. Since we could not consider source side imperfections, we could not consider Trojan Horse or other light injection attacks on our system \cite{Curr_s_Lorenzo_2025, mizutani2021security}. Whether SPS-based QKD can be protected against such attacks remains an open question, both experimentally and theoretically. This work demonstrates that quantum dot-based single photon sources have the potential to be a feasible technology for use in quantum cryptographic tasks, building on work done in Ref. \cite{vajner2026coinflip}. Major points of improvement in our implementation are the stabilization of QBER, which is sensitive to temperature fluctuations in Alice's EOM. Furthermore, increasing the clock rate of the system from 80 MHz to the GHz range, as readily feasible with QD devices \cite{Rickert_2025,behrends2026}, will allow for substantial improvements in the key rates. 

\section{Acknowledgments}
The authors thank Norbert Lütkenhaus for helpful discussions and Jerome Wiesemann for comments on the manuscript. This work was funded by the Natural Sciences and Engineering Research Council of Canada (NSERC) Discovery Grant. It was partially conducted at the Institute for Quantum Computing, University of Waterloo, which is funded by the Government of Canada through Innovation, Science and Economic Development Canada (ISED). Author A.C. is partially funded by the NSERC Alexander Graham Bell CGS-M, and part of the work was conducted while funded by the German Academic Exchange Service (DAAD) RISE Scholarship. Authors Z.W. and D.T. are partially funded by the Mike and Ophelia Lazaridis Fellowship. The authors further acknowledge financial support by the German Federal Ministry of Research, Technology and Space (BMFTR) via the project “QuSecure” (Grant No. 13N14876) within the funding program Photonic Research Germany, the BMFTR joint projects “tubLAN Q.0” (Grant No. 16KISQ087K), "NetiQueT" (Grant No. 16KIS1912), as well as QuNET+ICLink (Grant No. 16KIS1967) in the context of the federal government’s research framework in IT-security “Digital. Secure. Sovereign.”. H.L., S.L., H.N., and Z.N. acknowledge financial support by the Chinese Academy of Sciences Project for Young Scientists in Basic Research (Grant No. YSBR-112) and National Natural Science Foundation of China (Grant No. 12494601).

\section{Author Contributions}
A.C. and Z.W. contributed to the security analysis in this work under the supervision of D.T., K.K. and D.V. contributed to the experiment under the supervision of L.R., M.v.H. and T.H. L.R. designed and fabricated the single-photon source under supervision of T.H. and based on the quantum dot wafer material grown by H.L., S.L., H.N., and Z.N.;
\bibliography{references}
\appendix
\section{Entropic Uncertainty Relation Security Proof}\label{AppendixProof}
The security of QKD protocols depends on formalizing the idea that an eavesdropper is limited in how much information she can learn about the produced key \cite{tupkary2025qkdsecurityproofsdecoystate}. To do so, it is necessary to specify our model of BB84, and to note relevant quantities. This appendix treats all the steps involved in proving security for the protocol in the presence of device imperfections and can be broken down into the following : 
\begin{enumerate}
    \item In \href{sec:sourceReplaceAndEUR}{Section 1}, we begin by specifying the entropic uncertainty relation as well as the source replacement scheme, two integral components to proving security of our QKD implementation. We explain how these are used in proving QKD security.
    \item In \href{sec:newBounds}{Section 2}, we then recap work done in Ref. \cite{michaelPassiveEUR}, which provides statistical bounds on the required quantities for proving security. We further specify how these bounds change in the presence of imperfect single photon sources.
    \item In \href{sec:sourceMaps}{Section 3}, we elaborate on how to compute the needed multiphoton event bounds described in the above subsection by specifying source maps.
    \item In \href{sec:keyRateCalc}{Section 4}, we finally compute the final key rate and security parameters, as well as recap the appendix. 
\end{enumerate}

\subsection{Source Replacement Schemes and the Entropic Uncertainty Relation}\label{sec:sourceReplaceAndEUR}
\subsubsection{Source Replacement Scheme}
In order to prove the security of prepare and measure (P\&M) protocols, it is often more convenient to reformulate these as entanglement based protocols via the source replacement scheme \cite{PhysRevA.85.052310}. Instead of Alice preparing a state in every round and sending through an insecure quantum communication channel, Alice instead prepares an entangled state $\ket{\Psi}_{AA'}$, sends the $A'$ system to Bob through the usual quantum channel, and keeps the $A$ system to herself to measure later. The source-replaced states are chosen such that both scenarios are identical from Eve's perspective. Suppose Alice chooses the $\mathcal{Z}$ basis with probability $p_\mathcal{Z}^A$ where she sends the states $\ket{0}$ or $\ket{1}$ with equal probability, and the $\mathcal{X}$ basis with probability $p_\mathcal{X}^A$ where she sends the states $\ket{+}$ and $\ket{-}$ with equal probability, as is the case in the BB84 protocol. Then the joint state Alice prepares can be viewed as
\begin{align}
    \ket{\Psi}_{AA'}=\ket{\Psi^+}_{AA'}=\frac{1}{\sqrt{2}}\left(\ket{00}+\ket{11}\right)_{AA'},
\end{align}
with Alice's measurements being performed with the positive operator valued-measure (POVM) 
\begin{align}
    \Gamma^A_{\mathcal{Z},0} := p_\mathcal{Z}^A\ketbra{0},\ \Gamma^A_{\mathcal{Z},1} := p_\mathcal{Z}^A\ketbra{1},\ 
    \Gamma^A_{\mathcal{X},0} := p_\mathcal{X}^A\ketbra{+},\text{ and }\ 
    \Gamma^A_{\mathcal{X},1} := p_\mathcal{X}^A\ketbra{-}.
\end{align}
In the ideal case where Alice always prepares single photon states, the entire quantum phase of the protocol can be represented as Alice preparing $n_{1_A}$ of these in order to get some $\ket{\Psi}_{A^{\left(n_{1_A}\right)}A'^{\left(n_{1_A}\right)}}=\ket{\Psi}_{AA'}^{\otimes\left( n_{1_A}\right)}$ and sending all $n_{1_A}$ single photon $A'$ systems to Bob through some quantum channel, generating a joint entangled state state $\rho_{A^{\left(n_{1_A}\right)}B^{\left(n_{1_A}\right)}}$ . Importantly, we allow Eve to hold any purification of the joint state in her attack. 

\subsubsection{Entropic Uncertainty Relation}
Once we have formalized the source replacement scheme, we can now describe the entropic uncertainty relation (EUR) (See Theorem 1 of Ref. \cite{tomamichel2011uncertainty} as well as Section 6.1. of Ref. \cite{tupkary2025qkdsecurityproofsdecoystate}). In its most general form, we consider a joint state $\rho_{A^nB^nE}$, such as the one arising from the source replacement scheme above. Let $Z_A^n$ be a classical register of the measurement outcomes if Alice measures every subsystem in the $\mathcal{Z}$ basis, and let $X_A^n$ be a register of the measurement outcomes if Alice measures subsystem in the $\mathcal{X}$ basis. Then the EUR states that 
\begin{align}
    H_\mathrm{min}^\varepsilon (Z_A^n|E)_{\mathcal{M}_{\mathcal{Z}}(\rho)} + H_\mathrm{max}^\varepsilon (X_A^n|B^n)_{\mathcal{M}_{\mathcal{X}}(\rho)}\geq n,\label{eq:eurOG}
\end{align}
where $H_{\mathrm{min(max)}}^\varepsilon$ is the conditional smoothed min(max)-entropy and $\mathcal{M}_{\mathcal{Z}(\mathcal{X})}$ is a measurement channel measuring Alice's subsystems in the $\mathcal{Z}(\mathcal{X})$ basis. This can be viewed as a generalization of the Heisenberg uncertainty principle: given a tripartite state on systems $ABE$, if $A$ and $B$ are highly correlated in one basis then $A$ and $E$ must be uncorrelated in the complementary basis, and vice-versa. Since the min-entropy gives a quantification of Eve's information about Alice's key register\footnote{The min-entropy quantifies Eve's best guess about the information stored in the post-measurement register $Z_A$. Since this register is used to generate key rounds between Alice and Bob, it is operationally meaningful that this quantity be large, as it means Eve is only weakly correlated with the key register. For more details, see Ref. \cite{tupkary2024phaseerrorrateestimation}, Theorem 4.}, bounds on the max-entropy (which can be computed from observations in the protocol) allow us to bound Eve's knowledge. 
During the run of the protocol, there are $n_{\mathcal{Z},1_A,1_B}$ and $n_{\mathcal{X},1_A,1_B}$ rounds in each basis where Alice sent a single photon, and Bob received a single photon. We restrict our attention to the rounds where Alice sent a single photon and Bob received a single photon only, since multiphoton events are rare and the analysis of single photon rounds is greatly simplified. Because single-mode threshold detectors are blind to photon number coherence they are therefore equivalent to considering probabilistic photon states. In proving security, this means that without loss of generality we may assume Eve performs a QND measurement on the photon number output from Alice's lab to tailor her attack. As such, we can consider a well defined number of $n-$photon events in our analysis for $n=0,\ 1$ and $>1$. 
After compiling the the measurement rounds where the bases agree, Alice and Bob have a well defined notion of QBER, which we refer to as the error rates $e_{\mathcal{Z},1_A,1_B}$ and $e_{\mathcal{X},1_A,1_B}$. While not an actual measurement, Alice and Bob can also imagine what the QBER would have been had they measured all of their $\mathcal{Z}$ basis rounds in the $\mathcal{X}$ basis. This leads to the notion of \textit{phase error rate} $e_{\mathcal{Z},1_A,1_B}^\mathrm{Ph.}$, which is precisely this error rate for these rounds had they both measured their states in the $\mathcal{X}$ basis. Let $Z_A^{n_{\mathcal{Z},1_A,1_B}}$ be a register of Alice's measurement outcomes for the $n_{\mathcal{Z},1_A,1_B}$ $\mathcal{Z}$ basis rounds, and let $X_A^{n_{\mathcal{Z},1_A,1_B}}$ be a register of Alice's measurement outcomes had she measured those same rounds in the $\mathcal{X}$ basis. Furthermore, 
since we do not measure the phase error rate, we need to account for the fact that we can only estimate it. If we let $\mathcal{B}^\mathrm{Error}(\vec{n}^\mathrm{obs})$ be an upper bound on the phase error rate computed from protocol observations which holds with probability $1-\varepsilon^2$, i.e.
\begin{align}
    \mathrm{Pr}\left[e_{\mathcal{Z},1_A,1_B}^\mathrm{Ph.}\geq \mathcal{B}^\mathrm{Error}(\vec{n}^\mathrm{obs})\right]\leq\varepsilon^2,
\end{align}
then we may rearrange Eq. \eqref{eq:eurOG} and make use of Lemma 6 in Ref. \cite{tupkary2024phaseerrorrateestimation} (see also \cite[Eq. 5]{tomamichel2012tight}) to obtain
\begin{align}
    H^\varepsilon_\mathrm{min}\left(Z_A^{n_{\mathcal{Z},1_A,1_B}} \big| E\right)_{\mathcal{M_Z}(\rho)} \geq n_{\mathcal{Z},1_A,1_B}\left(1-h\left(\mathcal{B}^\mathrm{Error}(\vec{n}^\mathrm{obs})\right)\right),\label{eq:eurForQKD}
\end{align}
where $h$ is the binary Shannon entropy. As a result, the problem of proving security and bounding Eve's information is reduced to statistical bounds on the phase error rate.

\subsection{Passive Protocol EUR Bounds}\label{sec:newBounds}
Clearly, in Eq. \eqref{eq:eurForQKD}, there are two quantities which must be estimated from the measured protocol quantities, namely $n_{\mathcal{Z},1_A,1_B}$ and $e_{\mathcal{Z},1_A,1_B}^\mathrm{Ph.}$. Statistical bounds must be rigorously proven in the presence of finite size effects, device imperfections and the possibility of receiving more than a single photon in Bob's detectors. These statistical bounds are computed in terms of quantities that arise during the run of the protocol. The core quantities to be measured or estimated during the run of the protocol are listed in Table \ref{tab:quantities}, where we note that bold font denotes a statistical random variable with some associated underlying probability distribution.
\begin{table}[H]
\centering
\fbox{
    \begin{tabular}{c|c}
\textbf{Random Variables} & \textbf{Symbol} \\
\hline
Conclusive $\mathcal{Z}$ basis clicks from single photons (Alice and Bob) & $\mathbf{n}_{\mathcal{Z},1_A,1_B}$ \\
Conclusive $\mathcal{Z}$ basis clicks from single photons (Alice) & $\mathbf{n}_{\mathcal{Z},1_A}$ \\
Multiclick events from single photons (Alice) & $\mathbf{n}_{\mathrm{mc},1_A}$ \\ 
$\mathcal{Z}$ bit error rate from single photons (Alice and Bob) & $\mathbf{e}_{\mathcal{Z},1_A,1_B}^\mathrm{Obs.}$\\
$\mathcal{Z}$ phase error rate from single photons (Alice and Bob) & $\mathbf{e}_{\mathcal{Z},1_A,1_B}^\mathrm{Ph.}$ \\ 
Conclusive $\mathcal{X}$ basis clicks from single photons (Alice) & $\mathbf{n}_{\mathcal{X},1_A}$ \\ 
Observed errors in $\mathcal{X}$ basis from single photons (Alice) & $\mathbf{N}_{\mathcal{X},1_A}^\mathrm{obs}$\\
\end{tabular}%
}
\caption{Random variables (r.v.'s) associated with core quantities to be measured or estimated in protocol used to bound single photon key contributions and phase error rate in Lemma \ref{lem:MichaelBounds}. The notation $1_A$ signifies Alice emitted a single photon, while the notation $1_B$ signifies Bob received a single photon. By conclusive, we mean that a singular click was received in a round where Alice and Bob agreed on a basis.}\label{tab:quantities}
\end{table}
If these quantities were observed directly, then one would be able to bound the secure key length by \cite{michaelPassiveEUR, tupkary2024phaseerrorrateestimation}
\begin{align}
    \ell &\geq  H^\varepsilon_\mathrm{min}\left(Z_A^{n_{\mathcal{Z},1_A,1_B}}|E\right) - \lambda_\mathrm{EC} ({e}_{\mathcal{Z},1_A,1_B}^\mathrm{Obs.}, {n}_{\mathcal{Z}}, {n}_{\mathcal{X}}, N_{\mathcal{X}}^\mathrm{Obs.}) -  2\log(1/\varepsilon_\mathrm{PA}) - \log(2/\varepsilon_\mathrm{EV})\notag
    \\
    &\geq n_{\mathcal{Z},1_A,1_B}\left(1-h\left(e_{\mathcal{Z},1_A,1_B}^\mathrm{Ph.}\right)\right) - \lambda_\mathrm{EC} ({e}_{\mathcal{Z},1_A,1_B}^\mathrm{Obs.}, {n}_{\mathcal{Z}}, {n}_{\mathcal{X}}, N_{\mathcal{X}}^\mathrm{Obs.}) -  2\log(1/\varepsilon_\mathrm{PA}) - \log(2/\varepsilon_\mathrm{EV}).
\end{align}
The phase error rate $e^{\mathrm{Ph.}}$ is fundamentally impossible to observe and must be estimated. If Alice is in possession of a perfect single photon source, then we only need to estimate the fraction of rounds in which Bob also received a single photon. Such bounds on the requisite quantities are formulated in the following lemma. 
\begin{lem}[Bounds on Phase Error Rate and Single Photon Detections with Passive Optics, Theorem 1 of \cite{michaelPassiveEUR}.]\label{lem:MichaelBounds} Given a passive linear optical BB84 setup as in \cref{fig:lab_setup}, and conditioned on Alice sending $n_{1_A}$ single photon rounds with perfect qubit states, then 
\begin{align}
    \mathrm{Pr}\left[\mathbf{n}_{\mathcal{Z},1_A,1_B}\leq \mathcal{B}_{q_Z}^\mathrm{Single}(\mathbf{n}_{\mathcal{Z},1_A},\mathbf{n}_{\mathrm{mc},1_A})\ \vee \mathbf{e}_{\mathcal{Z},1_A,1_B}^\mathrm{Ph.}\geq\mathcal{B}^\mathrm{Error}_{a,\delta,q_Z}(\mathbf{n}_{\mathcal{X},1_A},\mathbf{n}_{\mathcal{Z},1_A},\mathbf{N}_{\mathcal{X},1_A}^\mathrm{obs},\mathbf{n}_{\mathrm{mc},1_A})\right]\leq \varepsilon^2_\mathrm{a}+\varepsilon^2_\mathrm{b}+\varepsilon_0^2 + \varepsilon_\mathrm{PNE}^2
\end{align}
where
\begin{align}
    \mathcal{B}^\mathrm{Single}_{q_Z}(\mathbf{n}_{\mathcal{Z},1_A},\mathbf{n}_{\mathrm{mc},1_A})&:=\mathbf{n}_{\mathcal{Z},1_A}-n(q_Z+\sqrt{\log(1/\varepsilon_0)n}) +\frac{2\log \varepsilon_\mathrm{PNE}}{4\lambda_{\min}^2}\notag \\&\ - \frac{1}{\lambda_{\min}}\mathbf{n}_\mathrm{mc,1_A}-\frac{1}{2\lambda_{\min}^2}\sqrt{\log(\varepsilon_\mathrm{PNE})(\log(\varepsilon_\mathrm{PNE}) - 4\lambda_{\min}\mathbf{n}_{\mathrm{mc},1_A})}, \label{eq:bSingle}
    \\
    \lambda_\mathrm{min} &:= 2\eta_\mathrm{min} ^2\bar{s}(1-\bar{s});\ \bar{s} = \frac{1}{2}+\max\{|s_\mathrm{min}-1/2|, |s_\mathrm{max}-1/2|\}
\end{align}
for $\eta_\mathrm{min}$ the minimum efficiency across all detectors, $s_\mathrm{min/max}$ the minimum/maximum ranges for Bob's beamsplitting ratio with $\overline{s}=\frac{1}{2}(s_\mathrm{max}+s_\mathrm{min})$, and
\begin{align}
    \mathcal{B}^\mathrm{Error}_{a,\delta,q_Z}(\mathbf{n}_{\mathcal{X},1_A},\mathbf{n}_{\mathcal{Z},1_A}, \mathbf{N}_{\mathcal{X},1_A}^\mathrm{obs},\mathbf{n}_{\mathrm{mc},1_A}) &= \frac{\mathbf{N}^\mathrm{obs}_{\mathcal{X},1_A}}{a\mathcal{B}_{q_Z}^\mathrm{Single}(\mathbf{n}_{\mathcal{Z},1_A},\mathbf{n}_{\mathrm{mc},1_A})}\notag
    \\
    &+\sqrt{-\frac{2\log(\varepsilon_\mathrm{a})}{a}\left(\frac{\mathbf{n}_{\mathcal{X},1_A}}{\mathcal{B}_{q_Z}^\mathrm{Single}(\mathbf{n}_{\mathcal{Z},1_A},\mathbf{n}_{\mathrm{mc},1_A})^2}+\frac{1}{\mathcal{B}_{q_Z}^\mathrm{Single}(\mathbf{n}_{\mathcal{Z},1_A},\mathbf{n}_{\mathrm{mc},1_A})}\right)}\notag
    \\
    &+\frac{\delta}{a}\left(\frac{\mathbf{n}_{\mathcal{X},1_A}}{\mathcal{B}_{q_Z}^\mathrm{Single}(\mathbf{n}_{\mathcal{Z},1_A},\mathbf{n}_{\mathrm{mc},1_A})}+1\right)\notag
    \\
    &+\sqrt{-2\log(\varepsilon_\mathrm{b})\left(\frac{\mathbf{n}_{\mathcal{X},1_A}}{\mathcal{B}_{q_Z}^\mathrm{Single}(\mathbf{n}_{\mathcal{Z},1_A},\mathbf{n}_{\mathrm{mc},1_A})^2}+\frac{1}{\mathcal{B}_{q_Z}^\mathrm{Single}(\mathbf{n}_{\mathcal{Z},1_A},\mathbf{n}_{\mathrm{mc},1_A})}\right)}.\label{eq:bError}
\end{align}
The values of $\varepsilon_i$ are free parameters relating the failure probabilities of the statistical bounds, while the values of $a, \delta$ and $q_Z$ are computed from the following device parameters that can be characterized before the experiment:
\begin{enumerate}
    \item $s_{\mathrm{min}/\mathrm{max}}$ the lower and upper bounds on the beamsplitting efficiency for Bob, 
    \item $\eta_{b,i}^{\mathrm{min}/\mathrm{max}},\ b\in\{\mathcal{Z},\mathcal{X}\},\ i\in\{0,1\}$ the lower and upper bounds on the (relative) detection efficiency of each individual detector for Bob,
    \item $d_{b,i}^{\mathrm{min}/\mathrm{max}},\ b\in\{\mathcal{Z},\mathcal{X}\},\ i\in\{0,1\}$ the lower and upper bounds dark count rates in each individual detector for Bob, and
    \item $p_{b}^A,\ b\in\{\mathcal{Z},\mathcal{X}\}$  the probability that Alice chooses to send a signal in the $\mathcal{Z}$ or $\mathcal{X}$ basis.
\end{enumerate}
\end{lem}
We refer to Ref. \cite{michaelPassiveEUR} for the exact expressions of $a, \delta$ and $q_Z$ (Eq.'s D40, D41 and E2, respectively). Broadly speaking, $a$ characterizes mismatch in basis selection, $\delta$ characterizes mismatch in detector efficiency and $q_Z$ characterizes dark count contributions to the final key rate.

In order to gain some intuition about these bounds, we can consider the perfect, asymptotic case. Here, we let $n\to\infty$ as well as set $s_\mathrm{min}=s_\mathrm{min}=1/2$, $\eta^{\mathrm{min}}_{b,i}=\eta^{\mathrm{max}}_{b,i}=1\ \forall b,i$, $d^{\mathrm{min}}_{b,i}=d^{\mathrm{max}}_{b,i}=1\ \forall b,i$ and $p_\mathcal{Z}=1/2$. In this case, one can compute that $\lambda_\mathrm{min} = 1/2$, and that $q_Z=0,\ a= 1,\ \delta = 0$. If, during the run of the protocol, one were to observe $n_{\mathcal{Z},1_A}\approx n_{\mathcal{X},1_A}$ and $n_\mathrm{mc}\ll n_{\mathcal{Z},1_A}$ (as can be reasonably expected an honest channel implementation), the bounds evaluate to
\begin{align}
    \frac{1}{n}\mathcal{B}_{q_Z}^\mathrm{Single} = \frac{1}{n}{n}_{\mathcal{Z},1_A} -\frac{1}{n}{n}_{\mathrm{mc},1_A}-\mathcal{O}\left(\frac{1}{n}
    \log(\varepsilon_\mathrm{PNE})\right)\approx \frac{1}{n}n_{\mathcal{Z},1_A}
\end{align}
and (picking $\varepsilon_a=\varepsilon_b$)
\begin{align}
    \frac{1}{n}\mathcal{B}_{a,\delta,q_Z}^\mathrm{Error} \approx \frac{1}{n}\frac{N_{\mathcal{X},1_A}^\mathrm{obs}}{n_{\mathcal{Z},1_A}/n}+\frac{2}{n}\sqrt{-2\log(\varepsilon_a)\left(\frac{n_{\mathcal{X},1_A}}{n_{\mathcal{Z},1_A}^2/n^2} + \frac{1}{n_{{\mathcal{Z},1_A}}/n}\right)}\approx \frac{N_{\mathcal{X},1_A}^\mathrm{obs}}{n_{\mathcal{Z},1_A}}.
\end{align}
Operationally, this means that with high probability (in this limited example) the majority of $\mathcal{Z}$ basis events stem from single photons and the phase error rate in the $\mathcal{Z}$ basis round is bounded by the (scaled) $\mathcal{X}$ basis error rate, as should be expected in the asymptotic, perfectly behaving and honest regime. The additional terms account for finite size fluctuations in the estimations as well as deviations from ideal devices.
 
Using these bounds, the computed key rate is $\varepsilon_\mathrm{sec} = 2\varepsilon_\mathrm{AT}+\varepsilon_\mathrm{EV}+\varepsilon_\mathrm{PA}$-secure, where $\varepsilon_\mathrm{AT}^2:= \varepsilon_0^2 + \varepsilon_\mathrm{PNE}^2 + \varepsilon_\mathrm{a}^2 + \varepsilon_\mathrm{b}^2$. 
Again note that when Alice makes use of a perfect single photon source, then the random variables $\mathbf{n}_{\mathcal{Z},1_A},\ \mathbf{n}_{\mathcal{X},1_A},\ \mathbf{n}_{\mathrm{mc},1_A}$ and $\mathbf{N}_{\mathcal{X},1_A}^\mathrm{obs}$ are directly observed, since Alice is always emitting single photons. Therefore, the above bounds are trivially computed by inputting the quantities measured during the protocol. Given that we do not have a perfect single photon source, these quantities must therefore be estimated themselves. We elaborate on this in the following subsection. 
\subsection{Imperfect source bounds and source maps}\label{sec:sourceMaps}
Lemma \ref{lem:MichaelBounds} can be easily applied if it is assumed Alice is in possession of a perfect single photon source, i.e. one that always emits single photons. In a practical QKD implementation, Alice does not have access to such a source but instead emits single photons with some non-unital probability, while emitting zero or multiple photons with other non-negligible probabilities. The core quantities to be estimated in Table \ref{tab:quantities} are of course $\mathbf{n}_{\mathcal{Z},1_A},\ \mathbf{n}_{\mathcal{X},1_A},\ \mathbf{n}_{\mathrm{mc},1_A}$ and $\mathbf{N}_{\mathcal{X},1_A}^\mathrm{obs}$, since these are no longer directly observed, but sampled from a larger distribution. To do so, it is convenient to replace Alice's source with a more convenient one during the security analysis through source map arguments.

\subsubsection{Source Maps}
A real single photon source can be modelled as probabilistically emitting zero, one or multiple photons with the probability distribution \cite{vajner2026coinflip}
\begin{align}
    \vec{p}_\mathrm{SPS} = \begin{pmatrix}
        p_{0_A} \\ p_{1_A} \\ p_{>1_A}
    \end{pmatrix} = \begin{pmatrix}
        1-\mu -\frac{1}{2}\mu^2 g^{(2)}(\tau=0)\\ \mu \\ \frac{1}{2}\mu^2 g^{(2)}(\tau=0)
    \end{pmatrix}\approx \begin{pmatrix}
        0.995\\0.005 \\ 2.875\cdot 10^{-7}
    \end{pmatrix},
\end{align}
where the mean photon number $\mu$ and the multiphoton suppression parameter $g^{(2)}(\tau=0)$ are to be measured experimentally (see Fig. \ref{fig:dotPerformance}) . It is helpful (for the purposes of theoretical analysis) to view this instead as Alice first preparing a nearly perfect single photon state (with small multiphoton contribution), followed by a modified attenuation channel 
\begin{align}
    \vec{p'}_\mathrm{SPS} \approx \begin{pmatrix}
        0 \\ 0.9999999713 \\ 2.875\cdot 10^{-7}
    \end{pmatrix}\mapsto \begin{pmatrix}
        0.995\\0.005 \\ 2.875\cdot 10^{-7}
    \end{pmatrix},
\end{align}
and performing the security analysis assuming the idealized probability distribution $\vec{p'}_\mathrm{SPS}$.   
This is an example of a \textit{source map} \cite{gottesman2004securityquantumkeydistribution,tupkary2025qkdsecurityproofsdecoystate}, a tool frequently used to simplify security proofs for imperfect sources. These source maps are defined as CPTP operations (quantum channels) on Alice's output state, which can be represented as block diagonal in Fock space \cite{Loredo_2019}, and acts on the block matrices as
\begin{align}
    \mathrm{CPTP}(A'\to A')\ni\Sigma : \begin{pmatrix}
        0 & & \\
         & p'_{1_A}\rho^{(1)}_{A'} & \\
         & & p_{>1_A} \rho^{(>1)}_{A'}
    \end{pmatrix} \mapsto\begin{pmatrix}
        p_{0_A} \rho^{(0)}_{A'} & & \\
         & p_{1_A}\rho^{(1)}_{A'} & \\
         & & p_{>1_A} \rho^{(>1)}_{A'}
    \end{pmatrix} ,
\end{align}
where each block $\rho_{A'}^{(n)}$ is itself a valid density operator. We can therefore imagine that Alice prepares a (nearly) perfect state, which then experiences some extra loss that only impacts the single photon subspace on the way to Bob. In order to then rigorously prove security security, the $\Sigma$ channel for each round is then absorbed into the overall quantum channel $\mathcal{E}_{(A')_1^n\to B_1^n}$, and therefore put under Eve's control, giving her strictly more power, and so not compromising the security of the protocol (For a pictorial view, see Fig. 1 of Ref. \cite{tupkary2025qkdsecurityproofsdecoystate} and for a rigorous mathematical treatment, see Lemma 8 of Ref. \cite{PRXQuantum.5.040315}). Furthermore, given that the state Alice prepares is block diagonal, there is a well defined number of single photon states that are created and sent during the run of the protocol, in which the shared state between Alice and Bob under the source replacement scheme can be viewed as the usual Bell $\Psi^+$ state \cite{michaelPassiveEUR}. During rounds in which Alice does not prepare a single photon state, Eve can perform a PNS attack which fully leaks these rounds' raw key bits. As such, we model the emitted multiphoton signals as orthogonal flag states to simplify our analysis, a technique also enabled by source maps \cite[Section 4.6.1]{tupkary2025qkdsecurityproofsdecoystate}. As such, if we can estimate the above random variables with high probability, then the task of proving security with the real single photon source emitting with distribution $\vec{p}_\mathrm{SPS}$ reduces to the task of proving security with the idealized source emitting with distribution $\vec{p'}_\mathrm{SPS}$\footnote{In reality of course, Alice emits mostly vacuum when $\mu$ is small, as is the case for us. This is accounted for in the observed statistics, and can be interpreted as the security proof "seeing" Alice's vacuum emissions as simply loss. It is also possible to account directly for these vacuum emissions directly with other statistical bounds, but these complicate the analysis. Code for this implementation in order to judge the differences in performance is available in the included GitHub repository.}.

\subsubsection{Accounting for Multiphoton Emissions}
Recall we want to estimate $\mathbf{n}_{\mathcal{Z},1_A},\ \mathbf{n}_{\mathcal{X},1_A},\ \mathbf{n}_{\mathrm{mc},1_A}$ and $\mathbf{N}_{\mathcal{X},1_A}^\mathrm{obs}$. We want a high probability lower bound on the first term and high probability upper bounds on the remaining terms, due to the random variables' roles in Eq.'s \ref{eq:bSingle} and \ref{eq:bError}. Since single photon versus multiphoton emissions are assumed to be binomially distributed in every round, we derive the following bounds. 
\begin{enumerate}
    \item (Lower bound on $\mathbf{n}_{\mathcal{Z},1_A}$) : In order to compute this bound, we first remark that $\mathbf{n}_{\mathcal{Z},1_A} = \mathbf{n}_\mathcal{Z} - \mathbf{n}_{\mathcal{Z},>1_A}$, so the problem can be reformulated as finding a high probability upper bound on $\mathbf{n}_{\mathcal{Z},>1_A}$. We do so in steps, and the first trivial bound is of the form $\mathrm{Pr}[\mathbf{n}_{\mathcal{Z},>1_A}\leq n_{>1_A}]=1$, which implies
    \begin{align}
        \sum_{n_{>1_A}}\mathrm{Pr}[\mathbf{n}_{>1_A}= n_{>1_A}]\mathrm{Pr}[\mathbf{n}_{\mathcal{Z},>1_A} \leq {n}_{>1_A}]=\mathrm{Pr}[\mathbf{n}_{\mathcal{Z},>1_A} \leq \mathbf{n}_{>1_A}]=1,
    \end{align}
    by definition of these random variables. Whereas $\mathbf{n}_{\mathcal{Z},>1_A}$ follows an unknown distribution due to Eve having full control over the quantum channel, the random variable $\mathbf{n}_{>1_A}$ follows a $\mathrm{Bin}(n;p_{>1_A})$ distribution (after $n$ emission rounds have been completed). As such, we can make use of common Clopper-Pearson intervals \cite{clopper1934use, szekely2000statistics}, which tells us that for any binomial random variable $\mathbf{X}\sim\mathrm{Bin}(n;p)$, $\mathrm{Pr}(\mathbf{X}\leq k) = 1- I_p(k+1,n-k)$ where $I_x(\alpha,\beta)$ is the regularized incomplete beta function with shape parameters $\alpha,\beta$ given by
    \begin{align}
        I_x(\alpha,\beta) :=\frac{1}{\int_0^1 \mathrm{d}t\ t^{\alpha-1}(1-t)^{\beta-1}} \int_0^x \mathrm{d}t\ t^{\alpha-1}(1-t)^{\beta-1}.
    \end{align}
    Since we want a high confidence upper bound for $\mathbf{n}_{>1_A}\sim \mathrm{Bin}(n;p_{>1_A})$, it follows we can set $J_{p_{>1_A},n}(k):=I_{p_{>1_A}}(k+1,n-k) = \varepsilon_{\mathcal{Z},1_A}^2$ and invert\footnote{The inversion of the regularized, incomplete beta function can be accomplished through numerical packages, such as in \texttt{MATLAB}.} to find 
    \begin{align}
        \mathrm{Pr}\left[\mathbf{n}_{>1_A}\leq J^{-1}_{p_{>1_A},n}(\varepsilon_{\mathcal{Z},1_A}^2)\right] \geq 1-\varepsilon_{\mathcal{Z},1_A}^2,
    \end{align}
    for some suitable numerical upper bound on $J^{-1}_{p_{>1_A},n}(\varepsilon_{\mathcal{Z},1_A})$. This is also trivially an upper bound on $\mathbf{n}_{\mathcal{Z},1_A}$, and by combining bounds we find
    \begin{align}
        \mathrm{Pr}\left[\mathbf{n}_{\mathcal{Z},1_A} \geq \mathbf{n}_\mathcal{Z} -J^{-1}_{p_{>1_A},n}(\varepsilon_{\mathcal{Z},1_A}^2) \right]\geq 1-\varepsilon_{\mathcal{Z},1_A}^2.
    \end{align}
    In particular, note that $J_{p,n}(k)=I_p(k+1,n-k)$, as defined by
    \begin{align}
        J_{p,n}(k) \propto \int_0^p \mathrm{d}t\ t^k (1-t)^{n-k-1},
    \end{align}
    is monotonically increasing in $p$, since $t^k (1-t)^{n-k-1}\geq 0\ \forall k\leq n$. Since monotonically increasing functions have monotonically increasing inverses, $J_{p,n}^{-1}$ is also monotonically increasing in $p$. If we have only characterized $p_{>1_A}$ in some range $\left[p_{>1_A}^\mathcal{L},p_{>1_A}^\mathcal{U}\right]$, a safe lower bound on the single photon $\mathcal{Z}$ basis rounds is then 
    \begin{align}
        \mathrm{Pr}\left[\mathbf{n}_{\mathcal{Z},1_A} \geq \mathbf{n}_\mathcal{Z} -J^{-1}_{p_{>1_A}^\mathcal{U},n}(\varepsilon_{\mathcal{Z},1_A}^2) \right]\geq 1-\varepsilon_{\mathcal{Z},1_A}^2.
    \end{align}
    \item (Upper bounds on $\mathbf{n}_{\mathcal{X},1_A}$, $\mathbf{n}_{\mathrm{mc},1_A}$ and $\mathbf{N}_{\mathcal{X},1_A}^\mathrm{obs}$) : Since we cannot make any type of statement on how little, if any, of the $\mathcal{X}$ basis rounds came from multiphoton emissions, we only get a trivial bound of the form
    \begin{align}
        \mathrm{Pr}[\mathbf{n}_{\mathcal{X}, 1_A}\leq n_{\mathcal{X}}]=1,
    \end{align}
    where $n_\mathcal{X}$ is the number of observed $\mathcal{X}$ basis events. It therefore follows that 
    \begin{align}
        \sum_{n_\mathcal{X}}\mathrm{Pr}[\mathbf{n}_\mathcal{X}=n_\mathcal{X}]\mathrm{Pr}[\mathbf{n}_{\mathcal{X}, 1_A}\leq n_{\mathcal{X}}]=\mathrm{Pr}[\mathbf{n}_{\mathcal{X}, 1_A}\leq \mathbf{n}_{\mathcal{X}}]=1.
    \end{align}
    The same applies for the number of multiclick events, giving us
    \begin{align}
        \mathrm{Pr}[\mathbf{n}_{\mathrm{mc}, 1_A}\leq \mathbf{n}_{\mathrm{mc}}]=1,
    \end{align}
    as well as for the observed $\mathcal{X}$ basis errors
    \begin{align}
        \mathrm{Pr}[\mathbf{N}_{\mathcal{X}, 1_A}^\mathrm{obs}\leq \mathbf{N}_{\mathcal{X}}^\mathrm{obs}]=1.
    \end{align}
    While these bounds seem pessimistic, we justify them by the fact that in an actual implementation of the protocol, most events will indeed be coming from single photon emissions, since the ratio of of single photons to multiphotons is on the order of $\frac{1}{2\mu g^{(2)}(\tau=0)}\approx 4000$.
\end{enumerate}
\subsection{Combining Bounds and Key Rate Calculation}\label{sec:keyRateCalc}
The combinations of the above bounds are trivial via a simple application of union bounds. We find that the updated bounds, which are now computable in terms of observed protocol quantities, are 
\begin{align}
    \mathrm{Pr}\bigg[&\mathbf{n}_{\mathcal{Z},1_A,1_B}\leq \mathcal{B}_{q_Z}^\mathrm{Single}\left(n_\mathcal{Z} -J^{-1}_{p_{>1_A}^\mathcal{U},n}(\varepsilon_{\mathcal{Z},1_A}^2),n_{\mathrm{mc},1_A}\right)\ \vee\ \notag\\
    &\mathbf{e}_{\mathcal{Z},1_A,1_B}^\mathrm{Ph.}\geq\mathcal{B}^\mathrm{Error}_{a,\delta,q_Z}\left(n_{\mathcal{X}},n_{\mathcal{Z}}-J^{-1}_{p_{>1_A}^\mathcal{U},n}(\varepsilon_{\mathcal{Z},1_A}^2),N_{\mathcal{X}}^\mathrm{obs},n_{\mathrm{mc}}\right)\bigg]&\leq\varepsilon^2_\mathrm{a}+\varepsilon^2_\mathrm{b}  + \varepsilon_0^2 + \varepsilon_\mathrm{PNE}^2 + \varepsilon_{\mathcal{Z},1_A}^2.  
\end{align}
Given these bounds, we can use Appendix B of Ref. \cite{michaelPassiveEUR} and Theorem 4 of Ref. \cite{tupkary2024phaseerrorrateestimation} to compute the final key rate. The variable length key rate is given by
\begin{align}
    \ell \geq
    \mathcal{B}_{q_Z}^\mathrm{Single}\left(n_\mathcal{Z} -J^{-1}_{p_{>1_A}^\mathcal{U},n}(\varepsilon_{\mathcal{Z},1_A}^2),n_{\mathrm{mc},1_A}\right) \left(1-h\left(\mathcal{B}^\mathrm{Error}_{a,\delta,q_Z}\left(n_{\mathcal{X}},n_{\mathcal{Z}}-J^{-1}_{p_{>1_A}^\mathcal{U},n}(\varepsilon_{\mathcal{Z},1_A}^2),N_{\mathcal{X}}^\mathrm{obs},n_{\mathrm{mc}}\right)\right)\right)\notag\\
    -\lambda_\mathrm{EC} (e_\mathcal{Z}^\mathrm{obs},e_\mathcal{X}^\mathrm{obs},n_{\mathcal{Z}},n_{\mathcal{X}}, n_\mathrm{mc}) - 2\log (1/2\varepsilon_\mathrm{PA}) - \log(2/\varepsilon_\mathrm{EV}),
\end{align}
where $e_\mathcal{Z}^\mathrm{obs}$ is a sample (chosen randomly with some small probability $p_\mathrm{Test}$) of the $\mathcal{Z}$ basis error rate revealed during testing. This key rate has security parameter
\begin{align}
    \varepsilon_\mathrm{AT}^2 := \varepsilon_0^2 + \varepsilon_\mathrm{PNE}^2 + \varepsilon_\mathrm{a}^2 + \varepsilon_\mathrm{b}^2 + \varepsilon_{\mathcal{Z},1_A}^2,
\end{align}
with each of the above $\varepsilon_i$ a free parameter that can be optimized. We again refer to Ref. \cite{michaelPassiveEUR} for explicit expressions of $a, \delta$ and $q_Z$.

\section{Blinking correction of employed QD emitter}\label{appendix:blinking}

\begin{figure}
    \centering
    \includegraphics[width=1\linewidth]{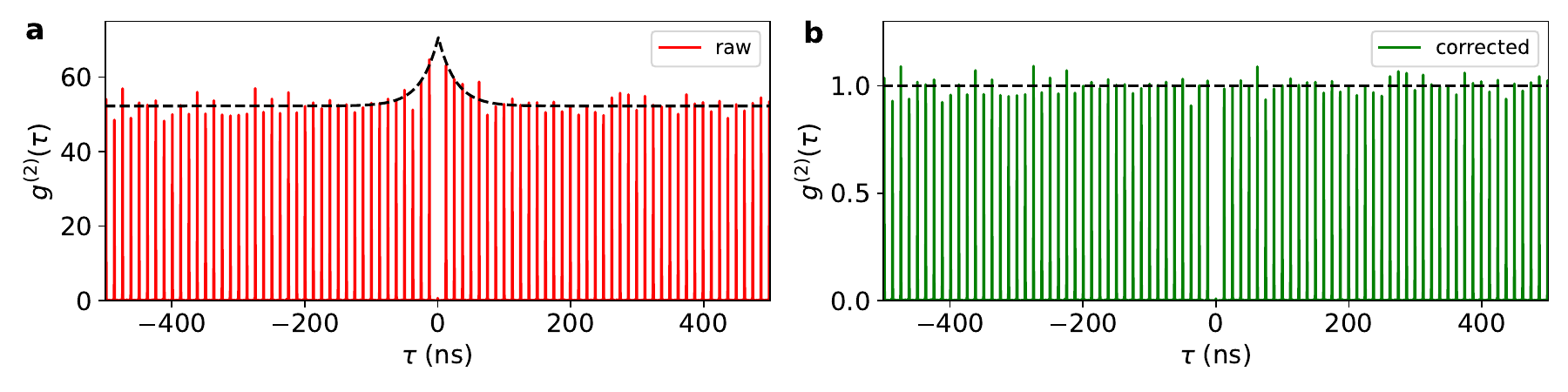}
    \caption{(a): Measured (raw) second order autocorrelation $g^{(2)}(\tau)$-histogram of the employed QD emitter. (b) Blinking-corrected $g^{(2)}(\tau)$-histogram by dividing the data in (a) by Eq.~\ref{eq:blinking_envelope} for the paramters listed in Tab~\ref{tab:blinking_fit}.}
    \label{fig:HBT_blinking}
\end{figure}

\noindent To account for blinking in the measured HBT histogram (see Fig.~\ref{fig:HBT_blinking}(a)), we fit the histogram peaks with the following blinking envelope function
\begin{equation}
g^{(2)}(\tau) = C_0+m \cdot \mathrm{e}^\frac{\vert\tau+\tau_0\vert}{\tau_\mathrm{blink}}
    \label{eq:blinking_envelope}
\end{equation}
and re-normalize the $g^{(2)}(\tau)$ by dividing the HBT histogram by Eq.~\ref{eq:blinking_envelope} with the fitted parameters in Table~\ref{tab:blinking_fit}. The resulting blinking-corrected HBT histogram is shown in Fig.~\ref{fig:HBT_blinking}(b), and $g^{(2)}(0)=0.017\pm0.001$ is obtained by integrated the remaining events in the time-bins around $\tau=0$ in a 12.5\,ns time-window, and dividing them by the averaged coindicences of the neighboring 80 peaks in their respective 12.5\,ns time-windows.

\begin{table}[H]
\centering
\fbox{
    \begin{tabular}{c|c}
\textbf{Fit parameter} & \textbf{Value} \\
\hline
$C_0$ & $52.2\pm0.3$ \\
$m$ & $18.5\pm3.7$ \\
$\tau_0$ & $(-0.90\pm2.72)$\,ns \\ 
$\tau_\mathrm{blink}$ & $(25.9\pm6.0)$\,ns \\
\end{tabular}%
}
\caption{Fitting parameters obtained from fitting the raw HBT histogram in Fig.~\ref{fig:HBT_blinking}(a) with the blinking envelope function in Eq.~\ref{eq:blinking_envelope}.}\label{tab:blinking_fit}
\end{table}

\end{document}